\documentclass[10pt]{article}
\usepackage[dvips]{epsfig}
\usepackage{epsf}
\newcommand{\beq}{\begin{equation}}

\newcommand{\eeq}{\end{equation}}
\newcommand{\bea}{\begin{eqnarray}}
\newcommand{\eea}{\end{eqnarray}}
\begin{document}

\begin{center}

{\large \bf Analysis of Intermediate-Energy Nucleus-Nucleus Spallation, 
Fission, and Fragmentation Reactions with the LAQGSM code
}\\

\vspace{0.5cm}

S.~G.~Mashnik$^1$, K.~K.~Gudima$^{2}$, R. E. Prael$^1$,  and A.~J.~Sierk$^1$

\vspace{0.2cm}
$^1${\em  
Los Alamos National Laboratory, Los Alamos, NM 87545, USA}\\

$^2${\em Institute of Applied Physics,
Academy of Science of Moldova, Kishinev, MD-2028, Moldova}\\
\end{center}

\vspace{0.3cm}
\begin{center}
{\bf Abstract}
\end{center}
{\noindent
The LAQGSM code has been recently developed at Los Alamos National 
Laboratory to simulate nuclear reactions for proton radiography
applications. We have benchmarked our code against most available 
measured data both for proton-nucleus and nucleus-nucleus interactions
at incident energies from 10 MeV to 800 GeV and have compared our results 
with predictions of other current models used by the nuclear community. 
Here, we present a brief description of our code and show illustrative 
results obtained with LAQGSM for neutron spectra measured recently by
Nakamura's groups for reactions induced by light and medium nuclei
on targets from $^{12}$C to $^{208}$Pb at several incident energies from 
95 to 
600 MeV/nucleon and with the recent GSI measurements of spallation,
fission, and fragmentation yields from A+p and  A+A reactions
at incident energies near and below 1 GeV/nucleon. Further necessary work 
is outlined.}

\vspace*{5mm}
{\noindent \bf \large Introduction}\\

During recent years, for a number of applications like 
Accelerator Transmutation of nuclear Waste (ATW),
Accelerator Production of Tritium (APT),
Rare Isotope Accelerator (RIA), Proton Radiography (Prad),
astrophysical work for NASA, and other projects,
we have developed at the Los Alamos National Laboratory
an improved version of the Cascade-Exciton Model (CEM), 
contained in the code CEM2k, to describe 
nucleon-, pion-, and photo-induced reactions at incident energies up to 
about 5 GeV \cite{CEM2k,CEM2kTsukuba}
and the Los Alamos version of the Quark-Gluon String Model,
realized in the high-energy code LAQGSM \cite{LAQGSM},
to describe both particle- and nucleus-induced reactions
at energies up to about 1 TeV/nucleon.

Both codes have been tested against most of the available data 
and compared with predictions of other modern codes
\cite{CEM2k}-\cite{Mashnik03c}. 
Our comparisons show that these codes describe 
a large variety of spallation, fission, and fragmentation reactions
quite reliably and often have a better predictive power than some
other available Monte-Carlo codes. 

In the present paper, we outline our models and show several 
typical results for nucleus-nucleus reactions
demonstrating that LAQGSM is a  reliable event generator that
can be used both in applications and in
fundamental nuclear research. 

Since LAQGSM uses modules
of CEM2k to describe the preequilibrium stages of nuclear
reactions and evaporation/fission of excited compound nuclei,
it is convenient for us to discuss both codes in this 
paper, although we show only results from LAQGSM.\\

{\noindent \bf \large CEM2k and LAQGSM Codes}\\

A detailed description of the initial version of the CEM may be found
in \cite{Gudima83}, therefore we outline here only its basic
assumptions.
The CEM assumes that reactions occur in three stages. The first
stage is the IntraNuclear Cascade (INC) 
in which primary particles can be re-scattered and produce secondary
particles several times prior to absorption by or escape from the nucleus.
The excited residual nucleus remaining after the 
cascade determines the particle-hole configuration that is
the starting point for the preequilibrium stage of the
reaction. The subsequent relaxation of the nuclear excitation is
treated in terms of an improved Modified Exciton Model (MEM) of preequilibrium 
decay followed by the equilibrium evaporative final stage of the reaction.
Generally, all three stages contribute to experimentally measured outcomes.

The improved cascade-exciton model in the code CEM2k differs from the 
older CEM95 version \cite{CEM95} by incorporating new 
approximations for the elementary cross sections used in the cascade,
using more precise values for nuclear masses and pairing energies, 
employing a corrected systematics for the level-density
parameters, 
adjusting the cross sections for pion absorption on quasi-deuteron 
pairs inside a nucleus, 
allowing for nuclear transparency of pions, including the Pauli principle 
in the preequilibrium calculation, and
improving the calculation of the fission widths.
Significant refinements and improvements in the 
algorithms used in many subroutines lead to a decrease of
computing time by up to a factor of 6 for heavy nuclei, which 
is very important when performing simulations with transport codes.
Essentially, CEM2k has a longer cascade stage,
less preequilibrium emission, and a longer evaporation stage
with a higher initial excitation energy, compared to its precursors
CEM97 \cite{CEM97} and CEM95 \cite{CEM95}.
Besides the changes to CEM97 and CEM95 mentioned above, we also made a 
number of other improvements and refinements, such as:
(i)
imposing momentum-energy conservation for each simulated event
(the Monte-Carlo algorithm previously used in CEM 
provided momentum-energy conservation only 
statistically, but not exactly for the cascade stage 
of each event),
(ii)
using real binding energies for nucleons at the cascade 
stage instead of the approximation of a constant
separation energy of 7 MeV used in previous versions of the CEM,
(iii)
using reduced masses of particles in the calculation of their
emission widths instead of using the approximation
of no recoil used previously, and
(iv)
a better approximation of the total reaction cross sections.
On the whole, this set of improvements leads to a much better description
of particle spectra and yields of residual nuclei and a better 
agreement with available data for a variety of reactions.
Details, examples, and further references may be found in
\cite{CEM2k,CEM2kTsukuba,Titarenko02}.

The Los Alamos version of the Quark-Gluon String Model
(LAQGSM) \cite{LAQGSM} is a further development of 
the Quark-Gluon String Model (QGSM) by Amelin, Gudima, and Toneev
(see \cite{Amelin90} and references therein) and is intended to describe
both particle- and nucleus-induced reactions at energies up to
about 1 TeV/nucleon. 
The core of the QGSM is built on a time-dependent version of the
intranuclear-cascade model developed at Dubna,
often referred in the literature simply
as the Dubna intranuclear Cascade Model (DCM) (see \cite{Toneev83}
and references therein).
The DCM models interactions of fast cascade particles (``participants")
with nucleon spectators of both the target and projectile nuclei and
includes interactions of two participants (cascade particles) as well.
It uses experimental cross sections (or those calculated by the Quark-Gluon 
String Model for energies above 4.5 GeV/nucleon) for these
elementary interactions to simulate angular and energy distributions
of cascade particles, also considering the Pauli exclusion
principle. When the cascade stage of a reaction is completed, QGSM uses the
coalescence model described in \cite{Toneev83}
to ``create" high-energy d, t, $^3$He, and $^4$He by
final-state interactions among emitted cascade nucleons outside 
of the colliding nuclei.
After calculating the coalescence stage of a reaction, QGSM
moves to the description of the last slow stages of the interaction,
namely to preequilibrium decay and evaporation, with a possible competition
of fission using the standard version of the CEM \cite{Gudima83}.
If the residual nuclei have atomic numbers 
with  $A \le 13$, QGSM uses the Fermi break-up model 
to calculate their further disintegration instead of using
the preequilibrium and evaporation models.
LAQGSM differs from QGSM by replacing the preequilibrium and
evaporation parts  of QGSM described according to the standard CEM 
\cite{Gudima83} with the new physics from CEM2k
\cite{CEM2k,CEM2kTsukuba} and has a number of improvements 
and refinements in the cascade and Fermi break-up models (in the
current version of LAQGSM, we use the Fermi break-up model only for
 $A \le 12$). A detailed description of LAQGSM and further
references may be found in \cite{LAQGSM}.

Originally, both CEM2k and LAQGSM were not able to describe fission reactions 
and production of light fragments heavier than $^4$He, as they had neither 
a high-energy-fission nor a fragmentation model.  Recently, we addressed 
these problems \cite{Mashnik02a,Mashnik02c} by further improving
our codes and by merging them with the Generalized Evaporation Model
code GEM2 developed by Furihata \cite{Furihata00,Furihata01}.

Our current versions of CEM2k and LAQGSM were
incorporated recently into the MARS \cite{MARS} and 
LAHET \cite{LAHET}
transport codes and are currently
being incorporated into MCNPX \cite{MCNPX}.
This will allow others to use our codes as event-generators
in these transport codes to simulate reactions with targets of
practically arbitrary geometry and nuclide composition.\\

{\noindent \bf \large Illustrative Results}\\

Recently, Nakamura's group measured neutron double-differential
cross sections from reactions induced by He, C, Al, and Ar
nuclei on C, Al, Cu, and Pb targets at several incident energies
from 95 to 600 MeV/nucleon (see \cite{Iwata01} and references therein).
We have calculated all these cross sections using LAQGSM.
As an example, Figure 1 shows our results for 
%interactions of 
560 MeV/nucleon $^{40}$Ar on 
%$^{12}$
C,
%$^{64}$
Cu, and 
%$^{208}$
Pb

%\newpage
\begin{figure}
\vspace{-5.cm}
\centerline{\hspace{-8mm} \epsfxsize 18cm \epsffile{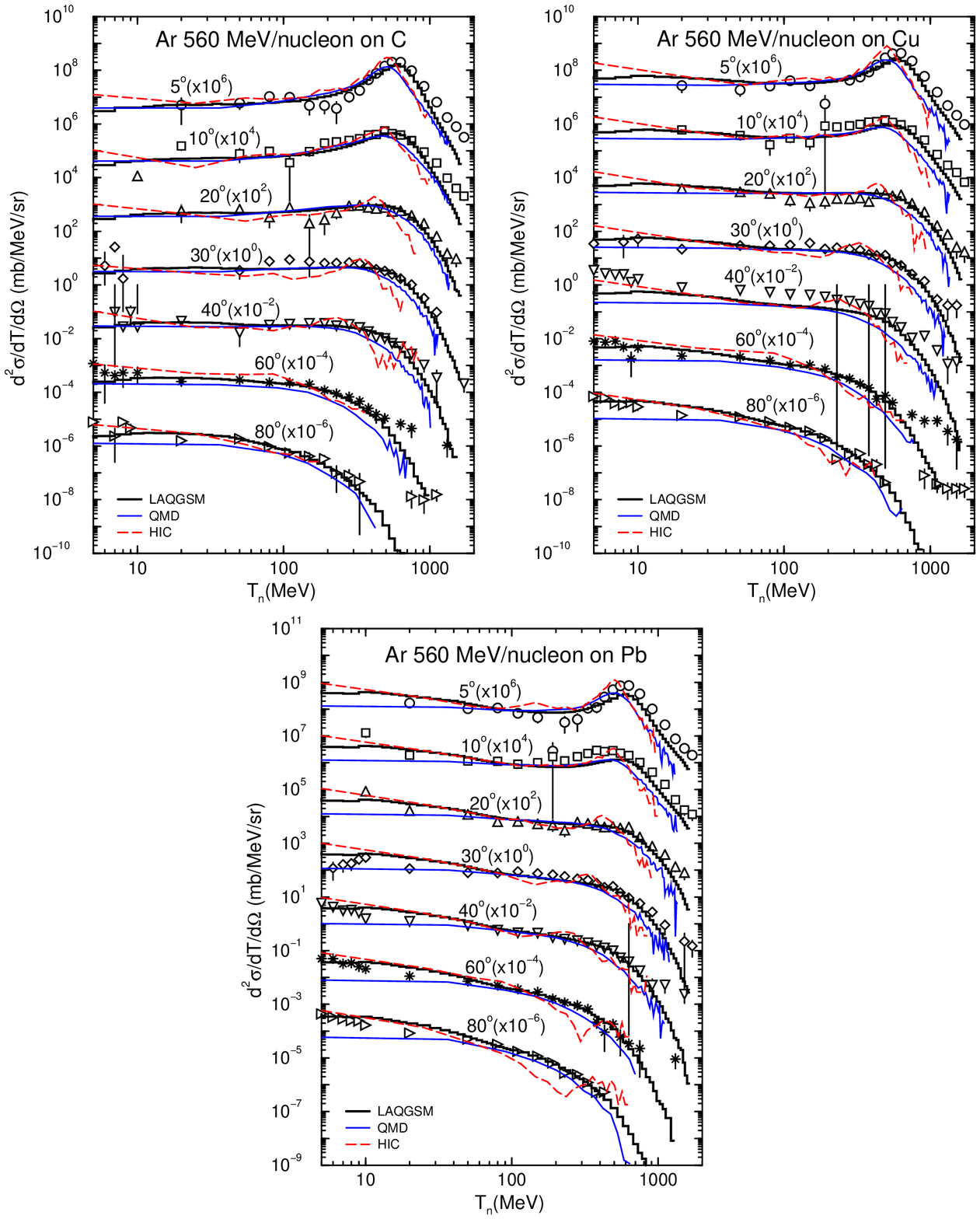}} 
\vspace{-27mm}
{\bf Figure 1.}
Comparison of measured \cite{Iwata01} double differential cross
sections of neutrons from 560 MeV/nucleon Ar beams on C, Cu and Pb 
with our LAQGSM results and calculations by QMD \cite{Aichelin91} 
and HIC \cite{Bertini74} from Iwata {\it et al.}
\cite{Iwata01}.
%\end{figure}

\vspace*{10mm}
\noindent{compared with experimental data and calculations
with the QMD \cite{Aichelin91} and HIC \cite{Bertini74}
models kindly provided to us by Nakamura's group.
We see that LAQGSM describes these data quite well and agrees with the 
measurements 
better than do QMD and HIC. Similar results are
obtained for all the other reactions measured by this group.
}

\end{figure}

%\newpage

Recently at GSI in Darmstadt, Germany, a large number of measurements 
have been performed using inverse kinematics
for interactions of $^{56}$Fe, $^{208}$Pb and $^{238}$U
at 1 GeV/nucleon and $^{197}$Au at 800 MeV/nucleon with 
liquid $^1$H. These measurements provide
a very rich set of cross sections for production of practically
all possible isotopes from such reactions in a ``pure" form,
{\it i.e.}, individual cross sections from a specific given bombarding isotope
(or target isotope, when considering reactions in the usual kinematics,
p + A). Such cross sections are much easier to compare to models than the 
``camouflaged" data from $\gamma$-spectrometry measurements. These 
are often obtained only for a natural composition of isotopes in a target
and are mainly for cumulative production, whereas measured cross sections
contain contributions not only from the direct production of
a given isotope, but also from all its decay-chain precursors. 
In addition, many reactions where a beam of light, medium,
or heavy ions with energy near to or below 1 GeV/nucleon interact with 
different nuclei, from the lightest, d, to the heaviest, $^{208}$Pb
were measured recently at GSI. References on these measurements and
many tabulated experimental cross sections may be found on the Web
page of Prof.\ Schmidt \cite{SchmidtWebPage}. 
We have analyzed with CEM2k and LAQGSM 
all measurements done at GSI of which we are aware, both for
proton-nucleus and nucleus-nucleus interactions. Some examples of
our CEM2k and LAQGSM results compared with the GSI
data and calculations by other current models
for proton-nucleus reactions may be found in
%\cite{CEM2k,CEM2kTsukuba,Mashnik02a,Mashnik02c,Mashnik02d,Titarenko02,
%Mashnik03,Mashnik03b,Mashnik03c}.
[1,2,4,6,7,9-12].
This paper is devoted to nucleus-nucleus reactions, but for 
completeness sake, we show in Fig.\ 2 just one example of LAQGSM
results for p+A interactions; namely,
spallation, fission, and fragmentation product yields from 
p(1 GeV) + $^{238}$U compared with the GSI data \cite{Taieb02,Bernas03}.
Similar results are obtained for all other p+A
reactions measured at GSI for which we could find data.

We performed our calculation
of this reaction in 2002, after the measured spallation product cross sections
were published in \cite{Taieb02}, and published our results in
the 2002 LANL Theoretical Division Report of Activity \cite{T&NW2002}.
The experimental data on fission and fragmentation products were published
only in 2003 \cite{Bernas03};
therefore the LAQGSM results for fission and
fragmentation products shown in the two upper panels of 
Fig.\ 2 are pure predictions; they agree amazingly well with the
experimental data.

We note that all the results shown in the figures of this paper were 
calculated within a single approach, without fitting any parameters of 
LAQGSM. 

Below we focus only on nucleus-nucleus reactions measured recently
at GSI, and we start our analysis with the lightest target, d, namely
with the reaction $^{238}$U(1 GeV/A) + d shown in Fig.\ 3.
One can see that LAQGSM merged with GEM2 (LAQGSM+GEM2)
describes quite well both the spallation and fission product
cross sections and agrees with most
of the GSI data with an accuracy of a factor of two or better.

Fig.\ 4 shows an example of a reaction on a heavier target, $^9$Be, 
namely the reaction 1 GeV/nucleon $^{86}$Kr + $^{9}$Be measured
by Voss \cite{Voss95}, compared with our LAQGSM+GEM2 results.
No fission mechanism is involved in this reaction and all the
measured products published in \cite{Voss95} and shown in this figure
are described by our code only via spallation. Although LAQGSM+GEM2
underestimates significantly the yields of neutron-rich Rb isotopes,
otherwise there is a good agreement between the calculations and data
for all the other measured cross sections.

Fig.\ 5 shows an example of a reaction on a heavier target, $^{27}$Al, 
namely the reaction 790 MeV/nucleon  $^{129}$Xe + $^{27}$Al measured
at GSI by Reinhold {\it et al.} \cite{Reinhold98} and compared with
LAQGSM+GEM2 results. Although both the projectile and target are heavier
than for the example shown in Fig.\ 4, LAQGSM+GEM2 describes all the
products from the reaction shown in Fig.\ 5 as well using only
spallation. A very good agreement between the data and calculations
may be seen for all measured cross sections, except for the neutron-rich
Cs isotopes, whose charge is bigger than that of initial Xe nuclei of the 
beam, being produced by picking up a proton from the Al target rather than
by spallation processes. The situation observed in Fig.\ 4
for the production of neutron-rich Rb isotopes involves the same process.

Finally, Fig.\ 6 shows a heavy-ion-induced reaction measured at GSI  
\cite{Junghans97,Junghans98}, namely the yields of measured spallation
products from the interaction of a 950 MeV/nucleon $^{238}$U beam 
with copper compared with our
results. LAQGSM+GEM2 describes most of these
data with an accuracy of a factor of two or better
(the fission and fragmentation products are not yet published
and we show here only the measured spallation yields,
though we calculated all the products from this reaction).

Fig.\ 7 show an example of several exotic reactions, namely 
fragmentation of secondary beams of neutron-rich unstable
$^{19,20,21}$O and stable $^{17,18}$O isotopes on $^{12}$C targets
at beam energies near
\newpage

\begin{figure}[t!]
\vspace{-5.cm}
\centerline{\hspace{-8mm} \epsfxsize 20cm \epsffile{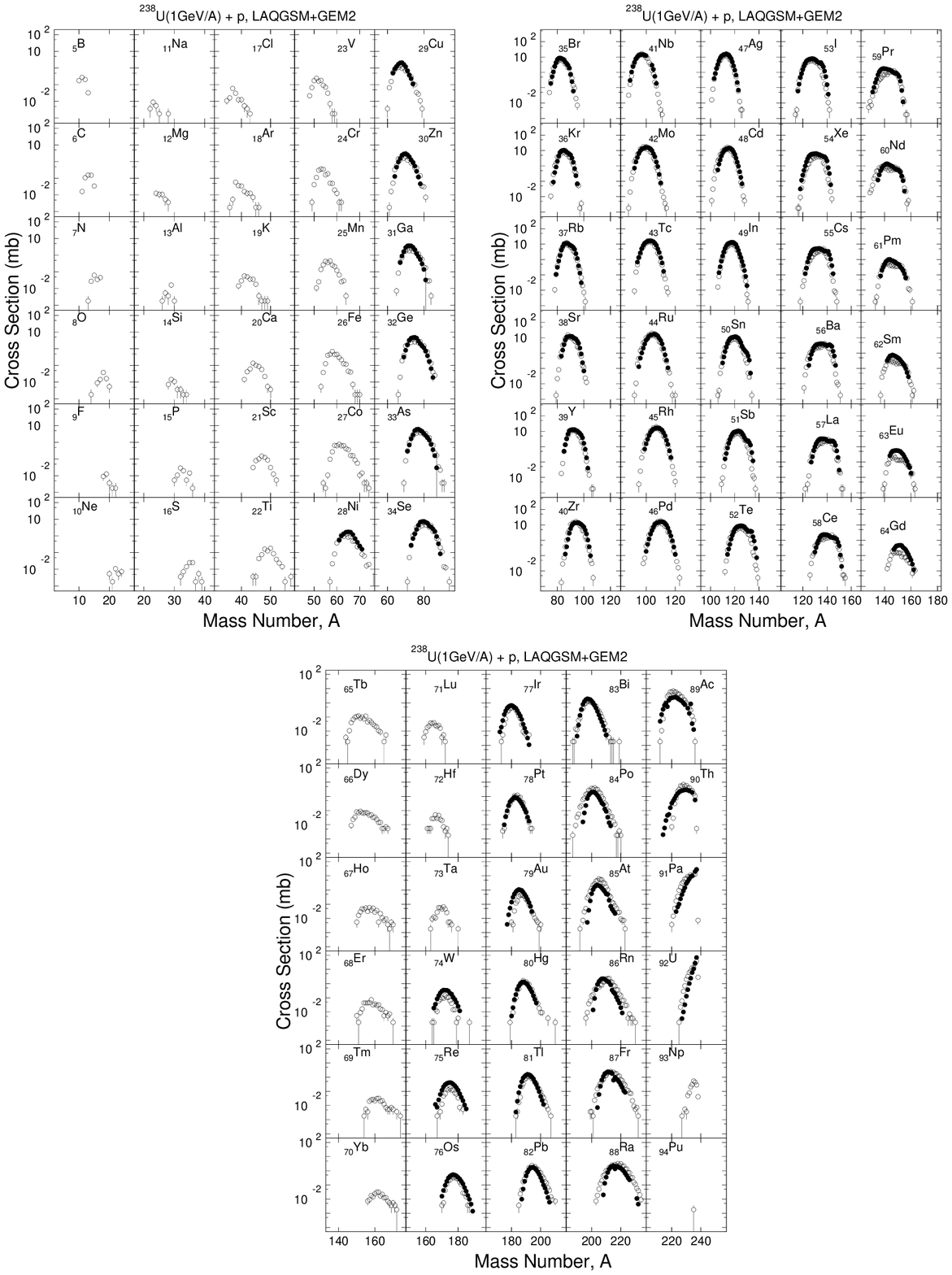}} 
\vspace{-3.0cm}
{\bf Figure 2.}
Comparison of measured \cite{Taieb02,Bernas03} 
spallation, fission, and fragmentation product cross
sections of the reaction $^{238}$U(1 GeV/A) + p (filled circles)
with our LAQGSM+GEM2 results (open circles). Experimental data for
isotopes from B to Co and from Tb to Ta are not yet available
so we present here only our predictions.
\end{figure}

\newpage

\begin{figure}[t!]
\vspace{-4.0cm}
\centerline{\hspace{-6mm} \epsfxsize 20cm \epsffile{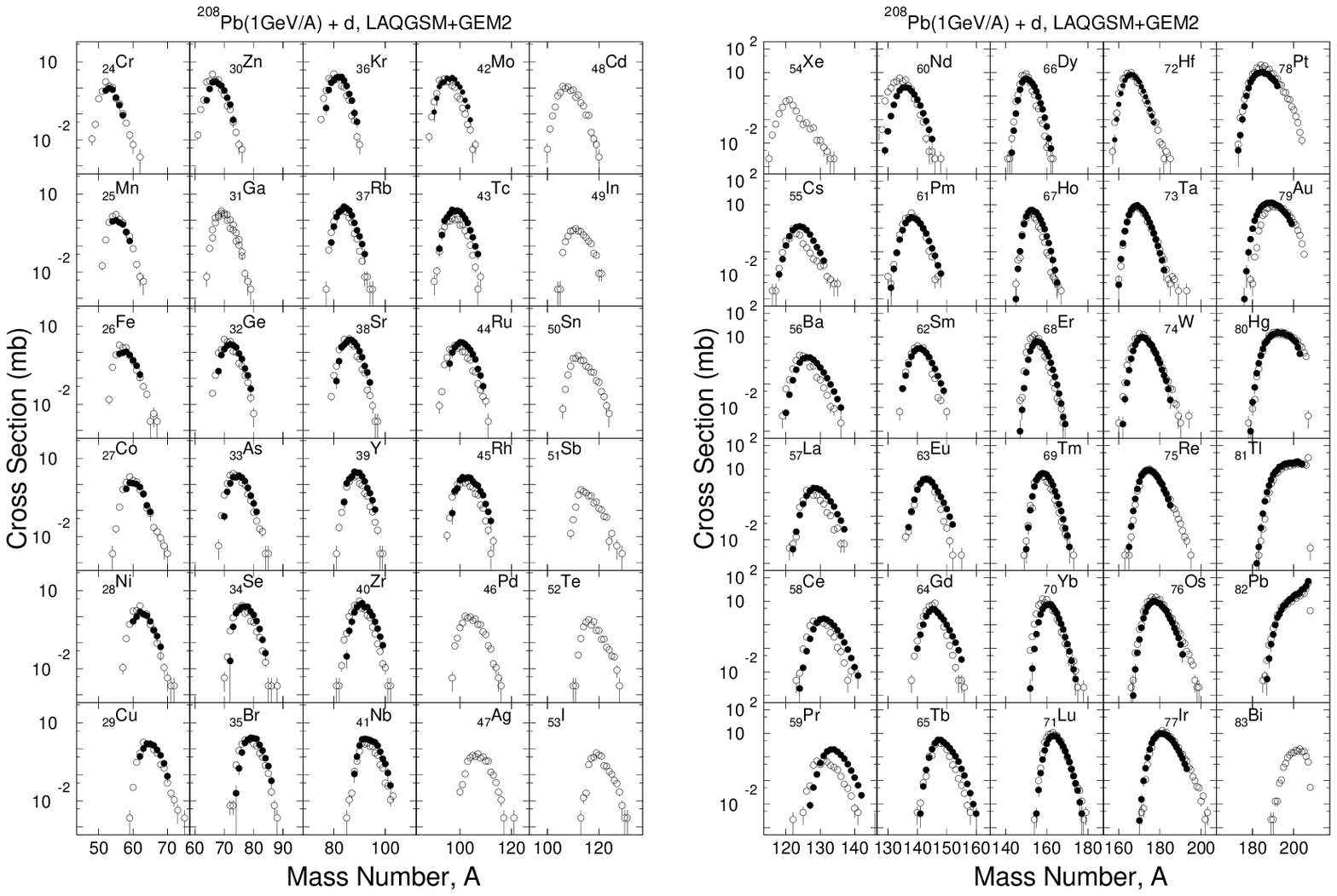}} 
\vspace{-14.0cm}
{\bf Figure 3.}
The same as in Fig.\ 2 but for the reaction
$^{208}$U(1 GeV/A) + d.
Experimental data (filled circles) are from \cite{Enqvist02};
open circles show our LAQGSM+GEM2 results.
%\end{figure}

%**************************** Begin Fig. 4 **************************
%\begin{figure}[h]
\begin{minipage}{8.0cm}
\vbox to 9.8cm {
\vspace*{-10mm}
\hspace*{-20mm}
\includegraphics[width=120mm,angle=-0]{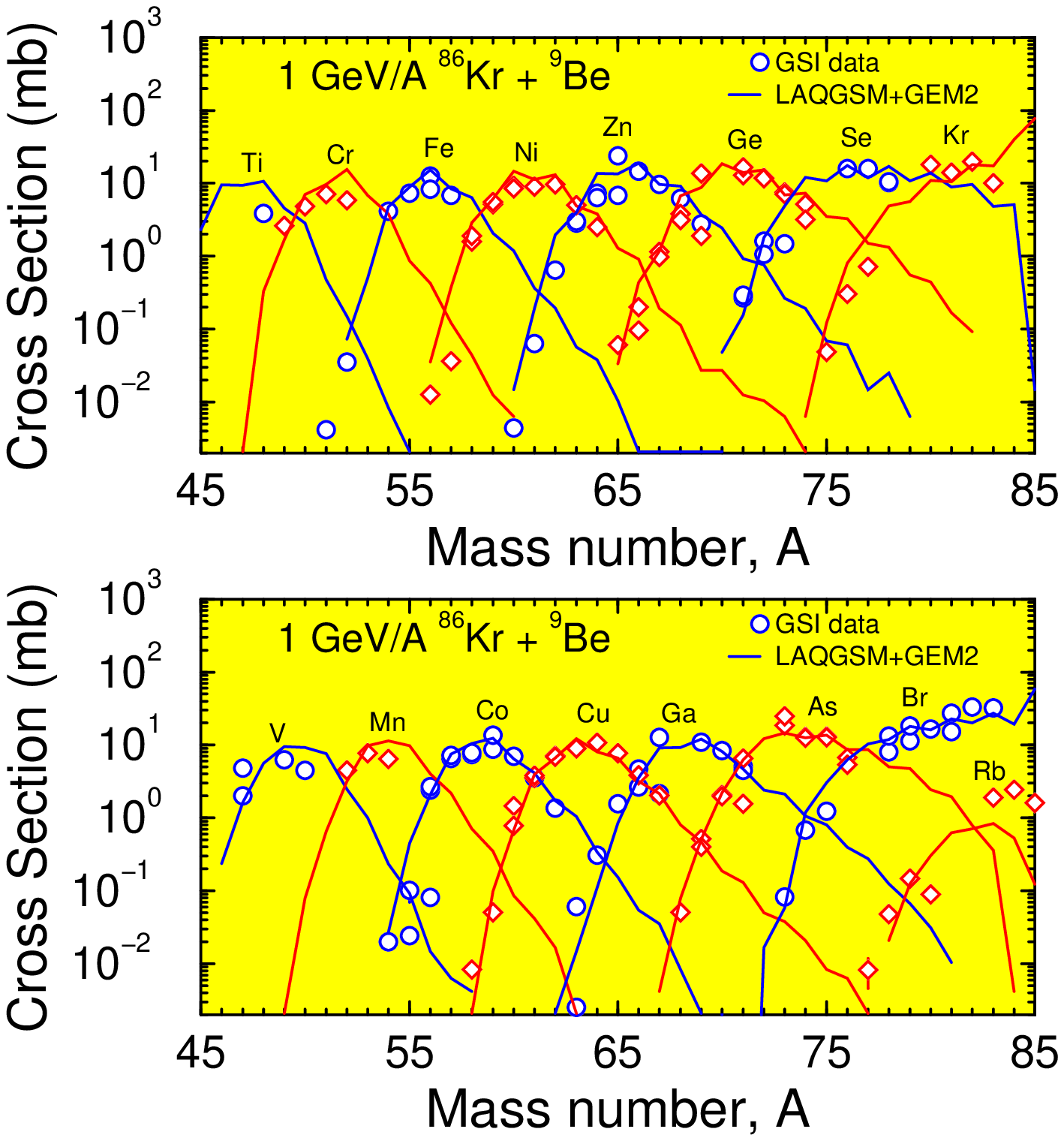}}
%\special{em:graph {e:/emtex/mytex/msp/mashnik2.msp}} }
\end{minipage}
\hfill
\begin{minipage}{7.0cm}
\vspace*{+5mm}
\begin{small}
{\bf Figure 4.}
Comparison of all measured \cite{Voss95}
cross sections of products from the reaction $^{86}$Kr + $^{9}$Be
at 1 GeV/nucleon (symbols) with our LAQGSM+GEM2 results (lines).\\
\\
\end{small}
\noindent{

 600 MeV/nucleon measured recently
at GSI \cite{Leistenschneider02}, compared with our LAQGSM+GEM2 results.
The secondary beams of $^{17-21}$O ions were produced in the fragmentation
of a primary $^{40}$Ar beam at 720 MeV/nucleon on a beryllium target
(see more details in \cite{Leistenschneider02}). The authors of this
measurement reproduced reasonably well the general trend
of their data with the empirical
parameterization EPAX \cite{EPAX} and with two versions of the
``abrasion-ablation" model \cite{Gaimard91,Carlson95}.
Nevertheless, the present version of the
EPAX parameterization does not contain any physical
description and does not reproduce the odd-odd effects in the production
cross sections.
}
\end{minipage}

%\vspace*{+2mm}
\vspace*{-1.5mm}
Both versions of the abrasion-ablation model \cite{Gaimard91,Carlson95}
do take into account even-even effects using experimental ground-state
masses and pairing shifts of $12 \sqrt{A}$ MeV, but apparently both
calculations overestimate the effect \cite{Leistenschneider02}.
We note that both EPAX \cite{EPAX}
and the abrasion-ablation model \cite{Gaimard91} failed to
reproduce well the recent GSI measurement of the
1 GeV/A $^{208}$Pb + Cu reaction \cite{deJong98}.
\end{figure}
%**************************** End Fig. 4 **************************

%\newpage

%**************************** Begin Fig. 5 **************************
\begin{figure}[h]
\begin{minipage}{11.0cm}
\vbox to 78mm {
\vspace*{-90mm}
\hspace*{-20mm}
\includegraphics[width=145mm,angle=-0]{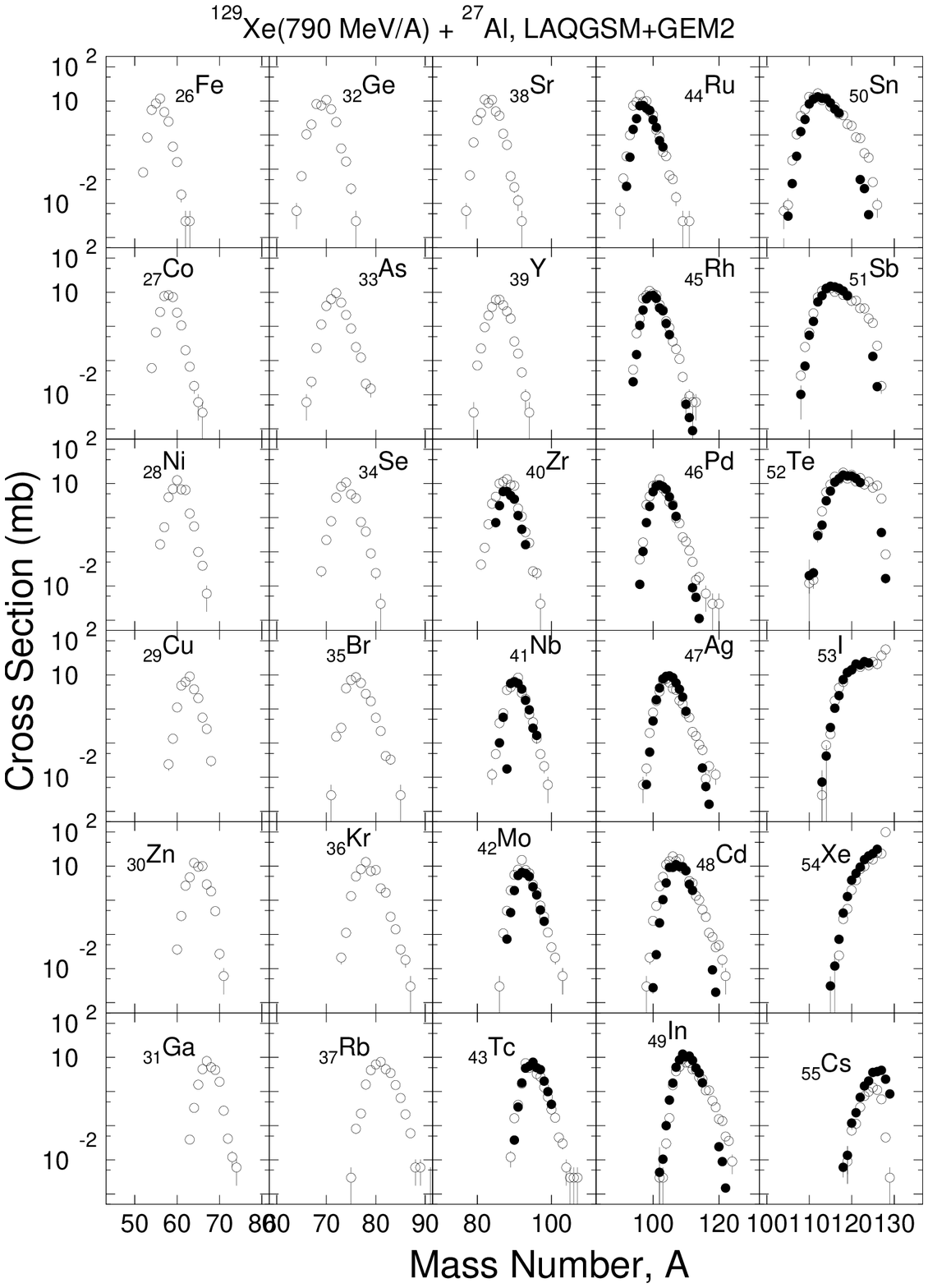}}
\end{minipage}
\hfill
\begin{minipage}{5.0cm}
\vspace*{-18mm}
\begin{small}
{\bf Figure 5.}
Comparison of all measured \cite{Reinhold98}
cross sections of products from the reaction $^{129}$Xe + $^{27}$Al
at 790 MeV/nucleon (filled circles) with our LAQGSM+GEM2 results 
(open circles). Isotopes from Fe to Y are not measured yet and we
present here only our predictions.\\
\\
\end{small}
\noindent{
This suggests we look first at the 1 GeV/A $^{208}$Pb + $^{64}$Cu
reaction  \cite{deJong98}
that gave problems to EPAX \cite{EPAX} and the abrasion-ablation
model \cite{Gaimard91} before trying to describe with LAQGSM+GEM2 the 
exotic measurements  \cite{Leistenschneider02} shown in Fig.\ 7.}
Our LAQGSM+GEM2 results for all cross sections measured by de Jong 
{\it et al.} \cite{deJong98} are compared with experimental data in
Fig.\ 8. One can see that LAQGSM+GEM2 describes reasonably well
all the measured data and we do not see any shifts either
to the neutron-rich or to the neutron-deficient regions observed
in \cite{deJong98} for EPAX and the abrasion-ablation model.
After addressing this reaction, we calculated with LAQGSM+GEM2
the reactions induced by neutron-rich $^{17-21}$O beams on $^{12}$C
targets measured in \cite{Leistenschneider02} and  shown in Fig.\ 7
as
\end{minipage}
\vspace*{0.5mm}
filled circles.
For completeness sake, we show in Fig.\ 7 calculated
cross sections for the production of all
O, N, and C isotopes, including the ones not measured in 
\cite{Leistenschneider02}, as well as yields of B and Be isotopes not 
measured at all, just as predictions. One can see that LAQGSM+GEM2
describes reasonably well all the measured cross sections,
and no worse than the abrasion-ablation model or phenomenological 
approximation EPAX do.  LAQGSM+GEM2 also predicts
significant yields for both neutron-rich and neutron-deficient
products not yet measured in  \cite{Leistenschneider02}.

\hspace{4mm}
In recent years, we observed in the literature an increased interest
in production and study of both neutron-rich and
neutron-deficient nuclei from different A+A reactions.
We analyzed some of these reactions with LAQGSM+GEM2.
One illustrative example is shown in Fig.\ 9, where we compare
the recent GSI measurement by Ozawa {\it et al.} 
\cite{Ozawa00} of the reaction 
$^{40}$Ar (1.05 GeV/nucleon) + $^9$Be with our results.
LAQGSM+GEM2 describes most of the measured neutron-rich product yields
quite well and reproduces correctly the change of the measured cross 
sections in an interval covering about six orders of magnitude. 
We believe that some of the overestimation by LAQGSM+GEM2 of the 
measured very neutron-rich product yields 
is related more to the limited statistics of our Monte-Carlo 
calculation (for the last measured neutron-rich nuclides with the
lowest cross sections, we have only one or two simulated events) 
than to some serious physics problems of our code. \\

{\noindent \bf \large Further Work} \\

\hspace{4mm}
From the results presented here and in the cited references,
we conclude that LAQGSM describes well (and
without any refitted parameters) a large variety of medium- and high-energy 
nuclear reactions induced both by nuclei and particles and is suitable for 
evaluations of nuclear data for applications and to study basic
problems in nuclear reaction science.
Merging our LAQGSM code with the Generalized 
\end{figure}
%**************************** End Fig. 5 **************************

\newpage
%**************************** Begin Fig. 6,7 **************************
\begin{figure}[h]
\begin{minipage}{11.0cm}
\vbox to 9.8cm {
\vspace*{-35mm}
\hspace*{-25mm}
\includegraphics[width=140mm,angle=-0]{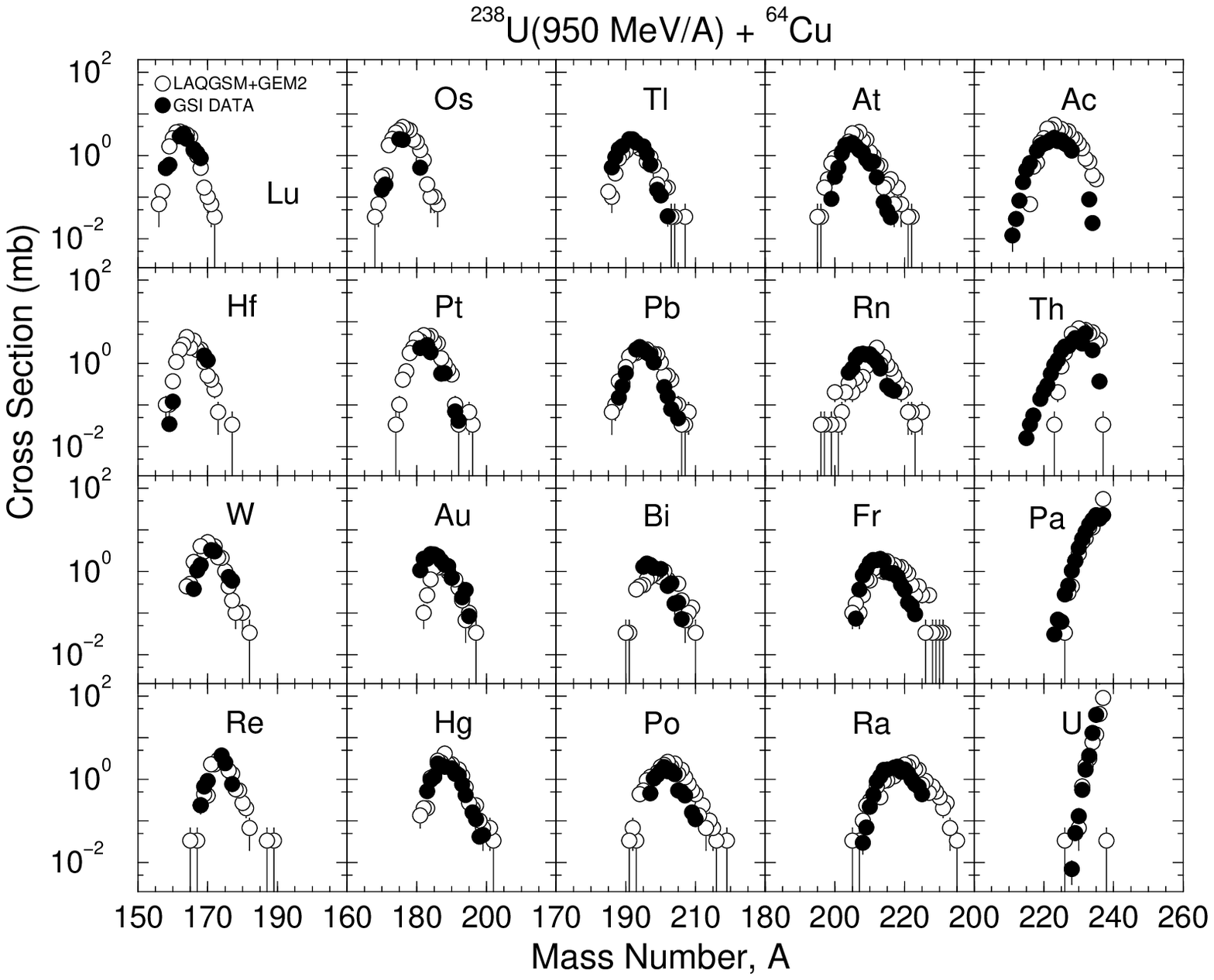}}
\end{minipage}
\hfill
\begin{minipage}{5.0cm}
\vspace*{-55mm}
\begin{small}
{\bf Figure 6.}
Comparison of all measured \cite{Junghans97,Junghans98}
cross sections of products from the reaction $^{238}$U + $^{64}$Cu
at 950 MeV/nucleon (filled circles) with our LAQGSM+GEM2 results 
(open circles).\\
\\
\end{small}
\noindent{

}
\end{minipage}
%\hfill
%\end{figure}
\begin{minipage}{11.0cm}
\vbox to 8.8cm {
\vspace*{-42mm}
\hspace*{-25mm}
\includegraphics[width=140mm,angle=-0]{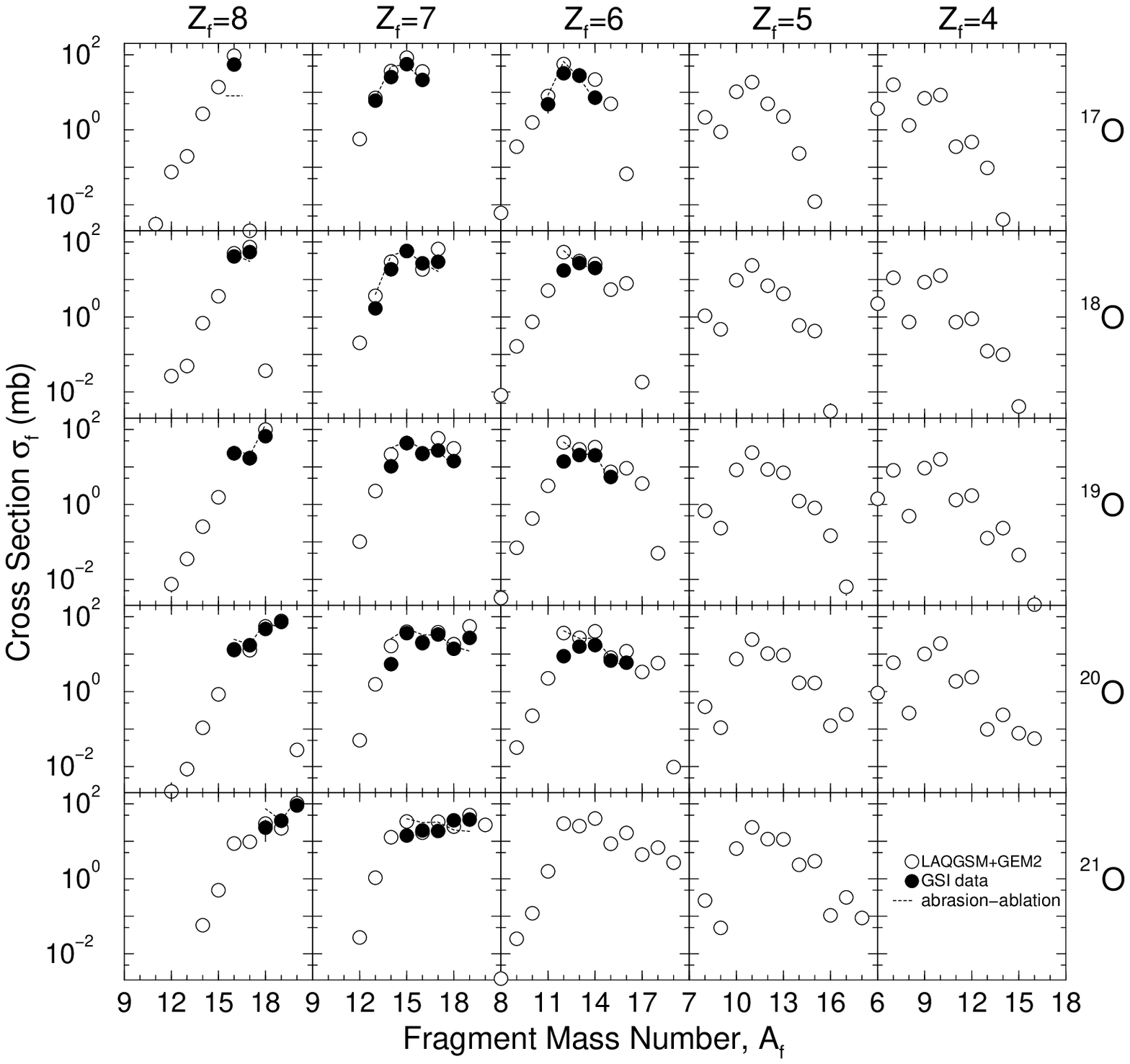}}
\end{minipage}
\hfill
\begin{minipage}{4.9cm}
\vspace*{-17mm}
\begin{small}
{\bf Figure 7.}
Cross sections of projectile fragments with nuclear charges $Z_f$
(shown on the top) and masses $A_f$ (shown on the bottom) produced
from $^{17-21}$O beams (shown on the right) in a $^{12}$C target.
Experimental data (filled circles) are from \cite{Leistenschneider02}.
Open circles show our LAQGSM+GEM2 results for the measured cross 
sections and predictions for several unmeasured isotopes.
Dashed lines show results by the the abrasion-ablation 
model \cite{Gaimard91} from \cite{Leistenschneider02}. \\
\end{small}
\\
\noindent{ 
Evaporation Model code GEM2 by Furihata \cite{Furihata00,Furihata01}
allows us to describe
reasonably well many fission and fragmentation reactions 
in addition to the spallation reactions already described 
well by LAQGSM. This does not means that LAQGSM+GEM2 is without
}

\end{minipage}
 problems.
For instance, it does not reproduce well the mass distributions for 
some fission-fragment elements from the reaction 1 GeV/A 
$^{238}$U + $^{208}$Pb measured recently at GSI \cite{Enqvist99},
although it still reproduces very well the
integral mass- and charge-distributions of all products.
We think that the main reasons for
this problem are the facts that the current version of LAQGSM
does not take into account electromagnetic-induced fission
\cite{Heinz03}, and because the GEM2 code by Furihata merged at
present with our LAQGSM does not consider at all the 
angular momentum of emitted particles, and of the compound nuclei.
Both these factors are especially important for reactions with heavy ions
and less important for reactions with light ions or protons;
this would explain why the code works well in the case of
reactions induced by particles and light and medium nuclei but
fails in the case of U+Pb. Besides the problem of angular momentum,
the current version of GEM2 has several more drawbacks related to
its lack of self-consistency (see details in \cite{Mashnik02a}).
We may choose to use a model similar to the GEM2 approach in 
\end{figure}
%**************************** End Fig. 6&7 **************************

\newpage
%**************************** Begin Figs. 8&9 **************************
\begin{figure}[h]
\begin{minipage}{11.0cm}
\vbox to 8.8cm {
\vspace*{-45mm}
\hspace*{-25mm}
\includegraphics[width=140mm,angle=-0]{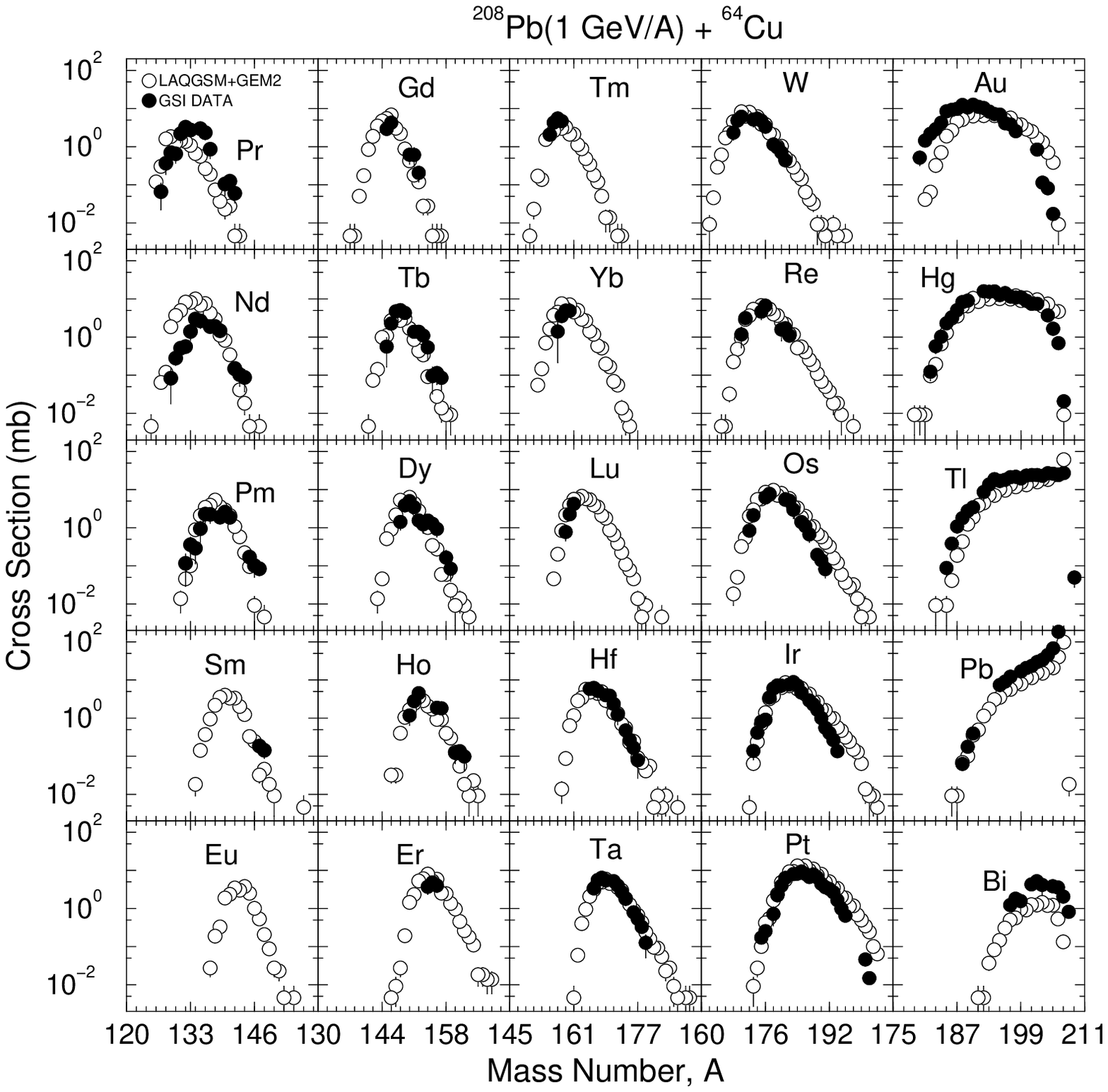}}
\end{minipage}
\hfill
\begin{minipage}{50mm}
\vspace*{-35mm}
\begin{small}
{\bf Figure 8.}
Comparison of all measured  \cite{deJong98}
cross sections of products from the reaction $^{208}$Pb + $^{64}$Cu
at 1 GeV/nucleon (filled circles) with our LAQGSM+GEM2 results 
(open circles).\\
\\
\end{small}
\end{minipage}

\begin{minipage}{70mm}
%\vspace*{-30mm}
%\hspace{-34mm}
\vbox to 80mm {
\vspace*{-35mm}
\hspace*{-60mm}
\includegraphics[width=180mm,angle=-0]{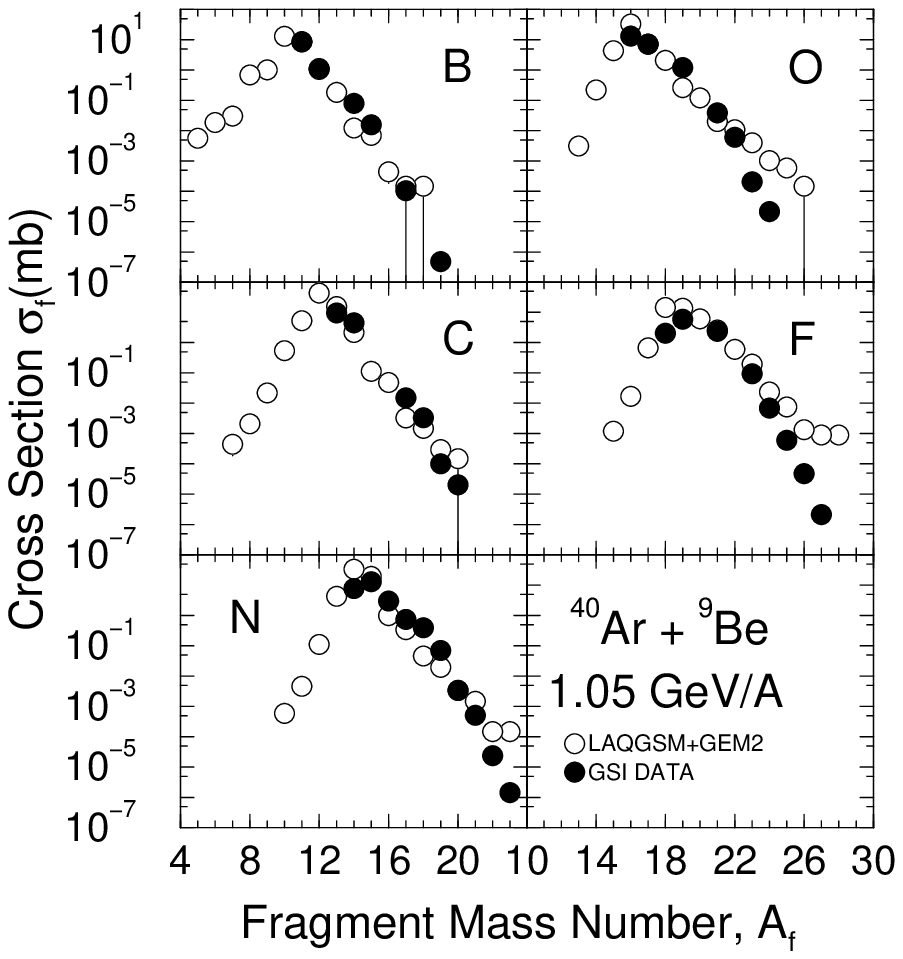}}
\end{minipage}
\hfill
\begin{minipage}{84mm}
\begin{small}
\vspace*{-1mm}
{\bf Figure 9.}
Experimental production cross sections \cite{Ozawa00}
for B to F isotopes from a 1.05 GeV/nucleon
 $^{40}$Ar beam on a $^9$Be target (filled
circles) compared with our LAQGSM+GEM2 results (open circles).\\
\\
\end{small}
\noindent{
the future versions
of our codes, but it must be significantly extended and further improved.
}
%\hspace{4mm} 
Our work on LAQGSM and CEM2k is not completed; we continue
their further development and improvement. 
Besides GEM2, we have investigated the well known
code GEMINI by Charity \cite{GEMINI} as an alternative way to describe
production of various fragments by merging GEMINI with both
LAQGSM and CEM2k, and we have also tested the
thermodynamical fission model by Stepanov \cite{Stepanov}
with its own parameterizations for mass and charge widths, 
level-density parameters, fission barriers, {\it etc.}, merging it
with both CEM2k and LAQGSM to describe fission.
In addition, we have started to extend CEM2k and LAQGSM and to
develop our own fission model, as briefly noted in \cite{Mashnik02c}.
The preliminary results we found for spallation, fission, and
fragmentation  products from several reactions we 

\end{minipage}

%\hspace{4mm} 
\vspace*{1mm}
tested so far using these approaches 
are very promising and we will present our results from these
studies in future papers.

\begin{center}
%\vspace{0.8cm}
{\it Acknowledgment}
\end{center}

\noindent
We thank 
Prof.\ Nakamura and Drs.\ Iwata and Iwase for sending us
numerical values of their measured neutron
spectra and results of calculations with QMD and HIC.
This study was supported by the U.\ S.\ Department of Energy and by the
Moldovan-U.\ S.\ Bilateral Grants Program, CRDF Projects MP2-3025 
and MP2-3045.
SGM acknowledge partial support from a NASA
Astrophysics Theory Program grant.
\end{figure}
%**************************** End Fig. 8&9 **************************

%\newpage

\begin{figure}

\end{figure}

\vspace*{-20mm}

\begin{thebibliography}{99}

\vspace*{-3mm}
\bibitem{CEM2k}
S. G. Mashnik and A. J. Sierk,
%``CEM2k---Recent Developments in CEM," 
{\it Proc. AccApp2000 (Washington DC, USA)}, p. 328
%La Grange Park, IL, USA,  2001 
(nucl-th/0011064). 
%see also

\vspace*{-3mm}
\bibitem{CEM2kTsukuba}
S. G. Mashnik and A. J. Sierk,
%``Recent Developments of the Cascade-Exciton Model of Nuclear Reactions,
%{\it Proc. ND2001 (Tsukuba, Japan)},  
J. Nucl. Sci. Techn. {\bf Supplement 2}, 720 (2002)
(nucl-th/0208074).

\vspace*{-3mm}  
\bibitem{LAQGSM}
K. K. Gudima, S. G. Mashnik, and A. J. Sierk,
``User Manual for the Code LAQGSM,"
Los Alamos National Laboratory Report LA-UR-01-6804, Los Alamos (2001).

\vspace*{-3mm}  
\bibitem{Titarenko02}
Yu. E. Titarenko {\it et al.},
%Cross sections for nuclide production in 1 GeV proton-irradiated Pb,
Phys. Rev. C {\bf 65}, 064610 (2002) (nucl-th/0011083).

\vspace*{-3mm}  
\bibitem{ Fertman02}
A. Fertman {\it et al.}, 
%I. Bakhmetjev, V. Batyaev, N. Borisenko, A. Cherkasov,
%A. Golubev, A. Kantsyrev, E. Karpikhin, A. Koldobsky, K. Lipatov,
%R. Mulambetov, S. Mulambetova, Yu. Nekrasov, M. Prokouronov,
%I. Roudskoy, B. Sharkov, Yu. Titarenko, V. Tutikov, V. Zhivun,
%G. Fehrenbacher, R.W. Hasse, D.H.H. Hoffmann, I. Hofmann, E. Mustafin,
%K. Weyrich, J. Wieser, {\bf S. Mashnik}, V. Barashenkov, and K. Gudima,
%``Induced Radioactivity Problem for High-Power Heavy-Ion Accelerator --
%Experimental Investigation and Longtime Predictions,''
%LANL Report LA-UR-02-3566 (2002),
%Proc. 14th Int. Symposium on Heavy Ion Fusion (HEF2002),
%Moscow, Russia, May 26-31, 2002;
%http://lib-www.lanl.gov/cgi-bin/getfile?00818851.pdf;
%E-print: nucl-ex/0209007;
{\it Proc. HEF2002},
Laser and Particle Beams {\bf 20}, 
%No. 3 
511 (2002) (nucl-ex/0209007).
% 511-514.

\vspace*{-3mm}
\bibitem{Mashnik02a}
S. G. Mashnik, K. K. Gudima, and A. J. Sierk,
%Merging the CEM2k and LAQGSM codes with GEM2 to describe
%fission and light-fragment production,
{\em  Proc. SATIF-6} 
%(SLAC, USA, 2002)}
%LANL Report LA-UR-02-0608;
%LANL Report LA-UR-03-2261,
% Los Alamos, 2002a
(nucl-th/0304012).

\vspace*{-3mm}
\bibitem{Mashnik02c}
S. G. Mashnik, A. J. Sierk, and K. K. Gudima,
%Complex-particle and light-fragment emission in the cascade-exciton model
%of nuclear reactions, 
{\em Proc. RPSD 2002 (Santa Fe, USA)}
%;  LANL Report LA-UR-02-5185, 2002c; 
(nucl-th/0208048).

\vspace*{-3mm}
\bibitem{Mashnik02b}
S. G. Mashnik, K. K. Gudima, N. V. Mokhov, R. E. Prael, and A. J. Sierk,
%Event Generator Benchmarking for Proton Radiography Applications,
{\em  Proc. SATIF-6} 
%(SLAC, USA, 2002)}
%; LANL Report LA-UR-02-0304; LANL Report LA-UR-03-1607, Los Alamos, 2002b; 
(nucl-th/0303041).

\vspace*{-3mm}
\bibitem{Mashnik02d}
S. G. Mashnik {\it et al.}, 
% R. E. Prael, A. J. Sierk, {\it et al.},
%V. F. Batyaev, S. V. Kvasova, R. D. Mulambetov, and Yu. E. Titarenko,
%Benchmarking Ten Codes Against the Recent GSI Measurements of the Nuclide
%Yields from $^{208}$Pb, $^{197}$Au, and $^{238}$U + p Reactions at
%1 GeV/nucleon,
%{\it Proc. ND2001 (Tsukuba, Japan)},  
J. Nucl. Sci. Techn. {\bf Supplement 2}, 785 (2002)
(nucl-th/0208075).

\vspace*{-3mm}
\bibitem{Mashnik03}   
S. G. Mashnik, K.~K.~Gudima, and R.~E.~ Prael,
%The CEM2k and LAQGSM Codes as Event Generators for RIA 
%Applications,
LANL Report LA-UR-03-0384, 
presented at {\it AccApp2003,
(San Diego, USA)}, to be published.

\vspace*{-3mm}
\bibitem{Mashnik03b}   
S. G. Mashnik, K. K. Gudima, R. E. Prael, and A. J. Sierk,
%Analysis of the GSI A+p and A+A Spallation, Fission, and Fragmentation
%Measurements with the LANL CEM2k and LAQGSM Codes (Abstract),
%LANL Report LA-UR-02-6959, 
%Los Alamos, 2002,
%to be presented at the 
{\it Proc. Workshop on Nuclear Data for the 
Transmutation of Nuclear Wastes, GSI, Germany, September 2003},
to be published.

\vspace*{-3mm}
\bibitem{Mashnik03c}  
S. G. Mashnik, K. K. Gudima, I. V. Moskalenko, R. E. Prael, and A. J. Sierk, 
%``CEM2k and LAQGSM as Event Generators for Space-Radiation-Shielding
%and Cosmic-Ray-Propagation Applications,''
{\it Proc. COSPAR 2002 (Houston, USA)},
%TX, USA, October 10--19, 2002;
%LA-UR-02-6558, Los Alamos (2002);
%http://lib-www.lanl.gov/cgi-bin/getfile?00783296.pdf;
%E-print: nucl-th/0210065, 2002;
to be published in {\em Advances in Space Research} (nucl-th/0210065).

\vspace*{-3mm}
\bibitem{Gudima83}   
K. K. Gudima, S. G. Mashnik, and V. D. Toneev,
%Cascade-exciton model of nuclear reactions,
Nucl. Phys. A {\bf 401}, 329 (1983). 

\vspace*{-3mm}
\bibitem{CEM95}
S.\ G.\ Mashnik, {\it User Manual for the Code CEM95}, JINR, Dubna, USSR;
OECD NEA Data Bank, Paris, France; 
http://www.nea.fr/abs/html/iaea1247.html;
RSIC-PSR-357, Oak Ridge, USA (1995).

\vspace*{-3mm}
\bibitem{CEM97}                 
S.~G.~Mashnik and A.~J.~Sierk, 
%``Improved Cascade-Exciton Model of Nuclear Reactions'', 
{\em Proc. 
%Fourth Int. Workshop on Simulating Accelerator Radiation Environments (
SARE-4 (Knoxville, USA, 1998)},
%edited by T. A. Gabriel, ORNL (1999)
p. 29 
(nucl-th/9812069).

\vspace*{-3mm}
\bibitem{Amelin90}
N. S. Amelin, K.~K.~Gudima, and V.~D.~Toneev,
%Further development of the model of quark-gluon strings for the
%description of high-energy collisions with a target nucleus,
Sov.~J.~Nucl.~Phys.~{\bf 52}, 172 (1990).
%[Yad.~Fiz.~{\bf 52} (1990) 272--282];

\vspace*{-3mm}
\bibitem{Toneev83}
V. D. Toneev and K.~K.~Gudima,
%Particle emission in light and heavy-ion reactions,
Nucl.~Phys.~A{\bf 400}, 173c (1983).

\vspace*{-3mm}
\bibitem{Furihata00}
S. Furihata, 
%Statistical analysis of light fragment production from medium energy
%proton-induced reactions,
Nucl.\ Instr.\ Meth.\ B{\bf 171}, 252 (2000).

\vspace*{-3mm}
\bibitem{Furihata01}
Shiori Furihata, 
{\em The Gem Code Version 2 Users Manual},
Mitsubishi Research Institute, Inc., Tokyo, Japan (2001);
Ph.D. thesis, Tohoku University (2003).

\vspace*{-3mm}  
\bibitem{MARS}
N. V. Mokhov,
%{\em The MARS Code System User's Guide},
Fermilab-FN-628 (1995); 
%more references and many details on MARS may be found at the Web page 
http://www-ap.fnal.gov/MARS/.

\vspace*{-3mm}  
\bibitem{LAHET}
R. E. Prael  and H.\ Lichtenstein, 
%{\em User Guide to LCS: The LAHET Code System},
LANL Report No.\ LA-UR-89-3014, Los Alamos (1989).
%http://www-xdiv.lanl.gov/XTM/lcs/lahet-doc.html.

\vspace*{-3mm}  
\bibitem{MCNPX}
{\it MCNPX$^{TM}$ User's Manual, Version 2.3.0}, 
edited by Laurie S. Waters,
LANL Report LA-UR-02-2607, Los Alamos (2002);
%see more references and many details at
http://mcnpx.lanl.gov/.
%{\it MCNPX$^{TM}$ User's Manual, Version 2.1.5},
%edited by Laurie S. Waters,
%Los Alamos National Laboratory Report LA-UR-99-6058, Los Alamos (1999).

\vspace*{-3mm}
\bibitem{Iwata01}  
Y. Iwata {\it et al.},
%T. Murakami, H. Sato, 
%H. Iwase, T. Nakamura, T. Kurosawa, L. Heibronn, R. M. Ronningen,
%K. Ieki, Y. Tozawa, and K. Niita,
%Double-differential cross sections for the neutron production from heavy-ion
%reactions at energies $E/A = 290-600$ MeV,
Phys. Rev. C {\bf 64}, 054609 (2001).

\vspace*{-3mm}
\bibitem{Aichelin91} 
J. Aichelin,
% ``Quantum" molecular dynamics -- a dynamical microscopic $n$-body
% approach to investigate fragment formation and nuclear equation of
% state in heavy ion collisions,
Phys. Rep. {\bf 202}, 233
%-360, 
(1991).
% Nos. 5 \& 6

\vspace*{-3mm}
\bibitem{Bertini74} 
H. W. Bertini {\it et al.},
% T. A. Gabriel, R. T. Santoro, 
% O. W. Hermann, N. M. Larson, and J. M. Hunt,
% {\em HIC-1: A First Approach to the Calculation of Heavy-Ion Reactions
% at Energies $\geq 50$ MeV/Nucleon},
Oak Ridge National Laboratory Report
ORNL-TM-4134, Oak Ridge (1974).

\vspace*{-3mm}
\bibitem{SchmidtWebPage}
K.-H. Schmidt, personal Web page, 2003:
http://www-wnt.gsi.de/kschmidt/.

\vspace*{-3mm}
\bibitem{Taieb02} 
J. Taieb {\it et al.},
% K.-H. Schmidt, L. Tassan-Got, P. Armbruster, J. Benlliure, M. Bernas,
% A. Boudard, E. Casarejos, S. Czajkowski, T. Enqvist, R. Legrain, S. Leray,
%B. Mustapha, M. Pravikoff, F. Rejmund, C. St\'ephan, C. Volant, and W. Wlazlo,
%``Measurement of Nuclide Cross-Sections of Spallation Residues in 
% 1 A GeV $^{238}$U + Proton Collisions,"
% Progress Report of HINDAS Work Package 6 (High- and Intermedate-Energy
% Nuclear Data for Accelerator-Driven Systems, 
HINDAS-9-02 Report (2002); 
hppt://www-wnt.gsi.de/kschmidt/Preprints/HINDAS-9-02/report8.pdf;
Nucl. Phys. A {\bf 724}, 413 (2003).

%\vspace*{-3mm}
%\bibitem{Taieb03}  % The same as {Taib02} but published in Nucl Phys. A, 2003
%J. Taieb {\it et al.},
% K.-H. Schmidt, L. Tassan-Got, P. Armbruster, J. Benlliure, M. Bernas,
% A. Boudard, E. Casarejos, S. Czajkowski, T. Enqvist, R. Legrain, S. Leray,
%B. Mustapha, M. Pravikoff, F. Rejmund, C. St\'ephan, C. Volant, and W. Wlazlo,
%``Evaporation Residues Produced in  Spallation Reaction
% $^{238}$U+p at 1 A GeV,"
% Nucl. Phys. A, {\bf 724}, No. 3-4, 413--430 (2003). 

\vspace*{-3mm}
\bibitem{Bernas03}
M. Bernas {\it et al.},
%  P. Armbruster, J. Benlliure, A. Boudard, E. Casarejos, S. Czajkowski, 
% T. Enqvist, R. Legrain, S. Leray, B. Mustapha, P. Napolitani, J. Pereira,
% F. Rejmund, M.-V. Ricciardi, K.-H. Schmidt,  C. St\'ephan, J. Taieb,
% L. Tassan-Got, and  C. Volant,
%``Fission-Residues Produced in the Spallation Reaction $^{238}$U + p at
% 1 A GeV,"
Preprint IPNO-DRE-2003-01/GSI 2003-11,
submitted to Nucl. Phys. A.
(nucl-ex/0304003).

\vspace*{-3mm}
\bibitem{T&NW2002}
%Stepan G. Mashnik and Arnold J. Sierk,
S. G. Mashnik and A. J. Sierk,
%``High-Energy Fission Reactions for Proton Rediography,"
pp. 30-31 in LANL Report
%{\it Theoretical Division Activities in Support of the Nuclear Weapons 
%Program, Winter 2002/2003},
LA-UR-03-0001, Los Alamos (2003).

\vspace*{-3mm}
\bibitem{Enqvist02} 
T. Enqvist {\it et al.}, 
% W. Wlazlo, P. Armbruster, J. Benlliure,
% M. Bernas, A. Boudard, S. Czajkowski, R. Legrain, S. Leray,
% B. Mustapha, M. Pravikoff, F. Rejmund, K.-H. Schmidt, 
% C. St\'ephan, J. Taieb, L. Tassan-Got, and C. Volant,
% ``Primary-residue Production Cross Sections and Kinetic Energy in
% 1$A$ GeV $^{208}$Pb + on Deuteron Reactions,"
Nucl. Phys. {\bf A703}, 435
%--465 
(2002).
%No. 1-2

\vspace*{-3mm}
\bibitem{Voss95}
B. Voss, 
% Berbd Voss, 
% ``Untersuchung der Projektilfragmentation und der Isotopenrennung 
% relativistscher Schwerionen am Fragmentseparator der GSI,"
% Institute f\"ur Kernphysik Technische Hochschule Darmstadt,
% Dortorarneit, M\"arz 1995 (240 pages)
Ph.D. thesis, KTH Darmstadt, 1995;
http://www-wnt.gsi.de/kschmidt/theses.htm. 

\vspace*{-3mm}
\bibitem{Reinhold98}
J. Reinhold {\it et al.},
% J. Friese, H.-J. K\"orner, R. Schneider, K. Zeitelhack, H. Geissel,
% A. Magel, G. M\"unzenberg, and K. S\"ummerer,
% ``Projectile Fragmentation of $^{129}$Xe at $E_{lab} = 790 A$ MeV,"
Phys. Rev. C {\bf 58}, 247 (1998).
% No. 1, pp. 247-255

\vspace*{-3mm}
\bibitem{Junghans97}
A. R. Junghans,
% Arnd  Rudolf Junghans, 
%{\em Untersuchungen zur Kollektivit\"at von Kernanregungen in der
%Fragmentation relativistscher Uranprojektile},
% {\em Investigation of Collectivity of Nuclear Excitations in the
% Fragmentation of Relativistic Uranium Projectiles},
Ph.D. thesis, Darmstadt TU, 1997; 
http://www-wnt.gsi.de/kschmidt/theses.htm. 

\vspace*{-3mm}
\bibitem{Junghans98}
A. R. Junghans {\it et al.},
%M.\ de Jong, H.-G.\ Clerc, et al., 
%A.\ V.\ Ignatyuk, G.\ A.\ Kudyaev, and K.-H.\ Schmidt,
% ``Projectile-fragment yields as a probe for the collective enhancement
% in the nuclear level density,"
Nucl. Phys. A {\bf 629}, 635
% -655, 
(1998).

\vspace*{-3mm}
\bibitem{Leistenschneider02}
A. Leistenschneider {\it et al.},
% T. Aumann, K. Boretzky, L. E. Canto, B. V. Carlson, D. Cortina, 
% U. Datta Pramanik, Th. W. Elze, H. Emling, H. Geissel, A. Gr\"unschloss,
% K. Helariutta, M. Hellstr\"om, M. S. Hussein, S. Ilievski, K. L. Jones,
% J. V. Kratz, R. Kulessa, Le Hong Khiem, E. Lubkiewicz, G. M\"unzenberg,
% R. Palit, P. Reiter, C. Scheidenberer, K.-H. Schmidt, H. Simon, 
% K. S\"ummerer, E. Wajda, and W. Wal\'us,
%``Fragmentation of Unstable Neutron-Rich Oxygen Beam,"
Phys. Rev. C {\bf 65}, 064607 (2002).
% No. 6 

\vspace*{-3mm}
\bibitem{EPAX}
K. S\"ummerer {\it et al.},
Phys. Rev. C {\bf 42}, 2546 (1990); K. S\"ummerer and B. Blank,
{\it ibid.} {\bf 61}, 034607 (2000).

\vspace*{-3mm}
\bibitem{Gaimard91}
J. J. Gaimard and K.-H. Schmidt,
Nucl. Phys. A {\bf 531}, 709 (1991).

\vspace*{-3mm}
\bibitem{Carlson95}
B. V. Carlson, M. S. Hussein, and R. C. Mastroleo,
Phys. Rev. C {\bf 46}, R30 (1992); B. V. Carlson,
{\it ibid.} {\bf 51}, 252 (1995).

\vspace*{-3mm}
\bibitem{deJong98}
M. de Jong {\it et al.},
Nucl. Phys. A {\bf 628}, 479 (1998).

\vspace*{-3mm}
\bibitem{Ozawa00}
A. Ozawa {\it et al.},
% O. Bochkarev, L. Chulkov, D. Cortina, H. Geissel, M. Hellstr\"om,
% M. Ivanov, R. Janik, K. Kimura, T. Kobayashi, A. A. Korsheninnikov,
% G. M\"unzenberg, F. Nickel, A. A. Ogloblin, M. Pf\"utzner, V. Pribora,
% H. Simon, B. Sit\'ar, P. Strmen, K. S\"ummere, T. Suzuki, 
% I. Tanihata, M. Winkler, and K. Yoshida,
% ``Production Cross-Section of Light Neutron-Rich nuclei from $^{40}$Ar
% Fragmentation at about 1 GeV/nucleon,"
Nucl Phys. A {\bf 673}, 411 (2000).
% No. 1--4, pp. 411--422

%%%%%%%%%%%%%%%%%%%%%%%%%%%%%%%%%%%%%%%%%% Our U + Pb problem %%%%%%%%%
\vspace*{-3mm}
\bibitem{Enqvist99}
T. Enqvist {\it et al.},
% J. Benlliure, F. Farget, K.-H. Schmidt, P. Armbruster, M. Bernas,
% L. Tissan-Got, A. Boudard, R. Legrain, C. Volant, C. B\"ockstiegel, 
% M. de Jong, and J. P. Dufour,
%``Systematics Experimental Survey on Projectile Fragmentation and Fission
% Induced in Collisions of $^{238}U at 1 GeV with Lead,"
Nucl. Phys. A {\bf 658}, 47 (1999).
% No. 1, pp. 47-66

%%%%%%%%%% Electromagnetic-induced fission, we do not take into account !!!! 
\vspace*{-3mm}
\bibitem{Heinz03}
A. Heinz {\it et al.},
% K.-H. Schmidt, A. R. Junhans, P. Armbruster, J. Benlliure, C. B\"ockstiegel,
% H.-G. Clerc, A. Grewe, M. de Jong, J. M\"uller, M. Pf\"utzner, 
% S. Steinh\"auser, and B. Voss,
%``Electromagnetic-Induced Fission of $^{238}$U Projectile Fragments, a
% Test Case for the Production of Spherical Super-Heavy Nuclei,"
Nucl. Phys. A {\bf 713}, 3 (2003).
% No. 1, pp. 2-23

\vspace*{-3mm}
\bibitem{GEMINI}
R.\ J.\ Charity {\it et al.}, %, R. J. and McMahan, M. A. and Wozniak, G. J. 
%and McDonald, R. J. and Moretto, L. G. and Sarantites, D. G. 
%and Sobotka, L. G. and Guarino, G. and Pantaleo, A. and
%Fiore, L. and Gobbi, A. and Hildenbrand, K. D.",
%``Systematics of Complex Fragment Emission in 
%Niobium-induced Reactions'',
Nucl. Phys. A {\bf 483}, 371 (1988);
%R. J. Charity,
%``$N-Z$ Distributions of Secondary Fragments and the
%Evaporation Attractor Line'', 
%{\it Phys. Rev.}{\bf C58}, 1073--1077 (1998);
%R. J. Charity {\it et al.},
% L. G. Sobotka, J. Cibor, K. Hagel, M. Murray, J.\ B.\ Natowitz, R.\ Wada,
% Y. El Masri, D. Fabris, G. Nebbia, G.\ Viesti, M.\ Cinausero, E.\ Fioretto, 
% G.\ Prete, A.\ Wagner, and H.\ Xu,
%``Emission of Unstable Clusters from Yb Compound Nuclei,''
%{\it Phys. Rev.} {\bf C63}, 024611 (2001);
http://wunmr.wustl.edu/~rc/.

\vspace*{-3mm}
\bibitem{Stepanov}
N. V. Stepanov,
%``Statistical Simulation of Excited Nuclei Fission. 
%1. Formulation of the Model'', 
ITEP Preprints 81 (1987) and 55-88 (1987), 
%Institute for Theoretical and Experimental Physics,
Moscow, USSR (1987).
%N. V. Stepanov,
%``Statistical Simulation of Excited Nuclei Fission. 
%2. Calculations and Comparison with Experiment,''
%ITEP Preprint 55-88 (1988),
%Institute for Theoretical and Experimental Physics,
%Moscow, USSR (1988).

\end{thebibliography}
\end{document}